# Modelling photothermal induced resonance microscopy: the role of interface thermal resistances

Francesco Rusconi[†], Marco Finazzi[†], John M. Stormonth-Darling[‡], Gordon Mills[‡], Michela Badioli[°], Leonetta Baldassarre[°], Michele Ortolani[°,§], Paolo Biagioni[†], and Valeria Giliberti[§,*]

[†] *Dipartimento di Fisica, Politecnico di Milano, Piazza Leonardo da Vinci 32, I-20133 Milano, Italy*

[‡] *Kelvin Nanotechnology Ltd, Rankine Building, Oakfield Ave, Glasgow, G12 8LT, UK*

[°] *Department of Physics, Sapienza University of Rome, Piazzale Aldo Moro 2, I-00185 Roma, Italy*

[§] *Istituto Italiano di Tecnologia, Center for Life Nano & Neuroscience, Viale Regina Elena 291, I-00161 Roma, Italy*

[*] E-mail: valeria.giliberti@iit.it

**Abstract**

Infrared (IR) nanospectroscopy by photothermal induced resonance (PTIR) is a novel experimental technique that combines the nanoscale resolution granted by atomic force microscopy (AFM) and the chemical labelling made possible by IR absorption spectroscopy. While the technique has developed enormously over the last decade from an experimental point of view, the theoretical modelling of the signal still varies significantly throughout the literature and misses a solid benchmark. Here, we report an analysis focused on the electromagnetic and thermal simulations of a PTIR experiment. Thanks to a control experiment where the signal is acquired as a function of the thickness of a polymer film and for different tip geometries, we find clear evidence that the interface thermal resistances play a key role in the determination of the measured signal and should therefore always be accounted for by any quantitative modelling.

**Introduction**

Photothermal induced resonance (PTIR) is a technique combining an atomic force microscope (AFM) measurement with infrared (IR) absorption spectroscopy that has been developed mainly over the last decade [1–13]. The employment of an AFM tip grants nanoscale resolution in the mapping of the sample, while the IR spectroscopy investigates the range of energies belonging to the fingerprint region, crucial for the chemical and conformational characterization of the sample. Applications range from nanoplasmonics [5,9,12] to subcellular imaging [3,7], from polymer blends [8,11] to protein membranes [13]. However, because of the novelty of the technique, a solid and widely accepted framework to model the experimental results is still lacking. In a typical PTIR experiment, IR light is absorbed by the sample, which consequently heats up and undergoes thermal expansion. The AFM tip measures such an expansion by means of an optical lever. Moreover, depending on the experimental conditions, the tip also grants a certain degree of confinement and enhancement of the IR electromagnetic field at its apex. By spanning the wavelength of the IR light, it is therefore possible to determine the absorption lines of the sample and thus obtain its molecular fingerprint spectrum with nanoscale resolution on the order of $\lambda/100$, i.e. well below the diffraction limit.

In the literature, different approaches have been applied to model the processes that determine the output of a PTIR experiment. One of the first theoretical analyses of the PTIR technique was published in 2010 by Dazzi *et al.* [1]. In this work, the authors study the different (mechanical, optical, and thermal) contributions to the PTIR signal. However, the investigated sample consists of a uniform sphere immersed in a homogeneous and isotropic medium, which represents a very different situation with respect to the film-on-substrate geometry often encountered in real situations. In particular, these two geometries highly differ from each other because of the drastically different boundary conditions, which are a crucial factor in the thermal analysis of the measurement. Lu *et al.*, in a paper from 2014, employ combined optical and thermal simulations together with a model based on the variation in the indentation depth to give an estimate of the force associated to the thermal expansion [6]. However, they study only molecular monolayers, which indeed constitute one of the most relevant and challenging situations from the point of view of PTIR applications but do not represent a solid benchmark to validate the modeling, since no thickness dependence is investigated. A different simplified model can be found in a more recent work by O'Callahan *et al.*, in which the thermal expansion force is estimated as the integral of the electric field intensity in the film [11]. However, this model does not take heat diffusion into account, making the assumption that the steady-state temperature increase has the same distribution as the electric field intensity. Recently, Chae *et al.* coupled a PTIR probe with photonic resonators to achieve high temporal resolution and directly measure the thermal conductivity of the sample [8]. In doing so, they highlight a specific thickness dependence that could be explained as due to the contribution of interface thermal resistances. However, they do not validate their analysis against optical and thermal numerical simulations.

In our work, starting from two sets of experimental data, we apply a modelling framework for a PTIR measurement with the help of optical and thermal simulations. By doing so, we provide evidence of the crucial role played by interface thermal resistances in the determination of the steady-state temperature in a typical experiment, a fact that must be taken into account for any successful modeling of the PTIR signal.

**Experimental setup and results**

The experiments that are the subject of this analysis were carried out using a nanoIR2 microscope (Anasys Instruments) equipped with a tunable quantum cascade laser emitting in the 5.8 µm to 6.3 µm wavelength range (Daylight Solutions). The IR light impinges on the sample with an intensity of about $10^7$ W m$^{-2}$ and an angle of incidence of 70° from the normal to the surface. The repetition rate of the 260 ns-long laser pulses matches the frequency of the second order bending mode of the cantilever (around 180 kHz), leading to an enhancement of the PTIR signal equal to the quality factor of the cantilever resonance [6]. The thermal expansion of the sample is then detected by demodulating the deflection signal of the cantilever, as measured by an independent optical lever. The sample under study is a 10-µm-long whisker of poly-methylmethacrylate (PMMA) with a nominal thickness varying between 0 and 200 nm and a width increasing from 200 to 1600 nm (**Figure 1a**). The whisker was fabricated by standard electron beam lithography over a gold-coated silicon substrate. The measurements were performed by probing the carbonyl line centered around 5.77 µm, which is one of the main vibrational frequencies of the PMMA and partially lays within the accessible spectral range of the laser. The PTIR spectrum is evaluated at different positions along the whisker, corresponding to different values of the thickness of the film, and the PTIR signal at each location is estimated by evaluating the spectral weight of the Lorentzian absorption peak after proper fitting.

We tested the behavior of two different tips (**Figure 1b-c**): a standard commercial gold-coated silicon tip and a custom-made triangular silicon nitride tip with a palladium patch at its end. The patch is about 1.95 µm long and it is designed to have its broad first-order dipole-antenna resonance at a wavelength of about 6 µm, which as discussed later on favors the confinement of the electric field at the apex [14,15] and increases the sensitivity of the tip to the sample surface. The two sets of measurements obtained with the two tips demonstrate qualitatively different behaviors: for the gold-coated tip we retrieve an almost linear dependence of the PTIR signal intensity on the PMMA thickness, while for the triangular tip we observe a sub-linear dependence (**Figure 1d-e**).

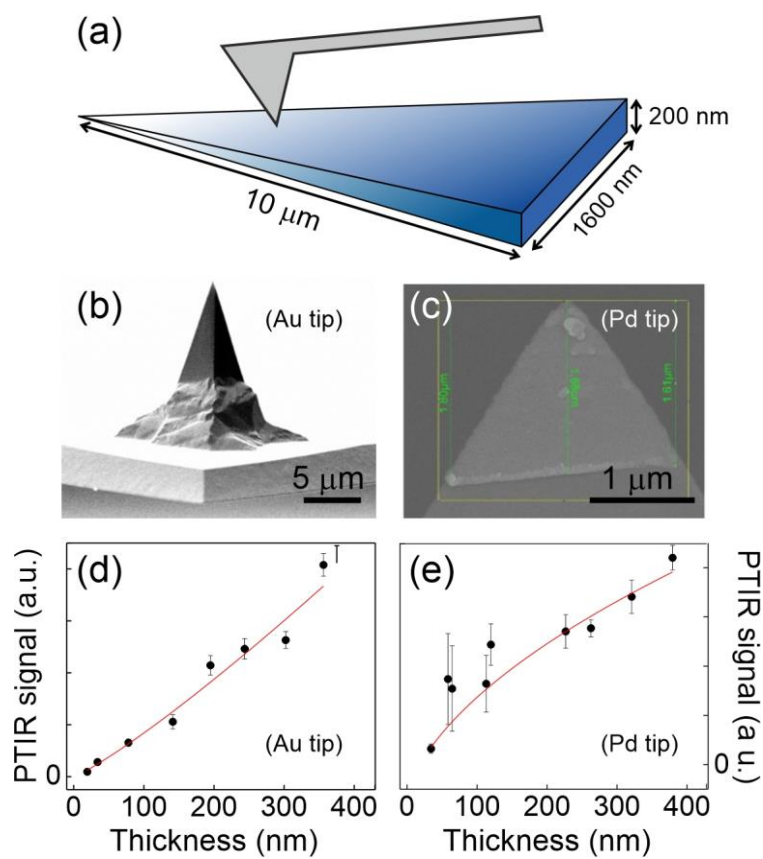

**Figure 1:** Details of the PTIR experiment. (a) Sketch of the geometry of the PMMA whisker. (b) Electron microscopy image of the Au tip employed in the experiment (side view). (c) Electron microscopy image of the Pd tip (top view of the resonant patch). (d) Experimental data points for the PTIR signal from the carbonyl line as a function of the thickness of the PMMA film with the Au tip. (e) Same as (d), obtained with the Pd tip. Error bars represent the standard deviation retrieved from the fitting procedure. Solid lines are a guide for the eye.

**Simulations and semi-analytical model**

The goal of this work is to use the experimental results described above as a benchmark to test the modeling of PTIR signals. In the experimental configuration employed for this work, the AFM tip is illuminated by a pulsed laser with a repetition rate matching the resonance frequency of the cantilever, therefore acting as a resonant driving force mediated by the periodic thermal expansions in the sample. In this situation, the oscillation amplitude (which the demodulated PTIR signal is proportional to) scales with the amplitude of the driving force. Therefore, the quantity investigated by the AFM tip is the elastic force arising from the photo-thermal expansion of the sample: $F = E\frac{\Delta z}{z_0}$, where $E$ is the Young modulus of the sample, $z_0$ its thickness, and $\Delta z$ is the thickness variation due to the thermal expansion, $z$ being the spatial coordinate perpendicular to the sample surface.

In order to estimate $\Delta z$ we perform optical and thermal simulations, which allow us to single out the crucial parameters for the modeling and analyze the regime under which we have a good correspondence between the simulations and the measurements. The workflow is divided into two steps: first we perform optical simulations, employing a finite-difference time-domain method (FDTD, Ansys/Lumerical) to evaluate the density distribution of the absorbed optical power. Then, we import the results into a second software that models heat diffusion employing a finite-element method (HEAT, Ansys/Lumerical) and treats the absorbed optical power density as the heat source, to eventually find the steady-state distribution of the temperature to which $\Delta z$ is related. Both the optical and the thermal simulations are performed in a two-dimensional environment: since we are dealing with a thin film, this is a reasonable approximation that greatly reduces the computational cost of the simulations. For the optical contribution, we perform two different sets of simulations for the two different tips. It is here worth noticing that the Pd patch is not in direct contact with the PMMA film, rather the underlying $Si_3N_4$ tip is. In order to replicate the analysis at different positions along the whisker, we run a set of simulations changing the value of the thickness of the film, from 10 to 320 nm, and adjusting the width accordingly as in the actual sample geometry. The illumination is provided by a plane wave source with the same intensity and angle of incidence as the experimental source. At the end of the optical simulations we obtain the electric field distribution inside the PMMA film (**Figure 2a-b**), from which the absorbed power density can be computed. This is then imported into the thermal software as a heat source for steady-state thermal simulations. The boundary conditions of the thermal simulations keep the temperature fixed at 300 K (room temperature) at the top and bottom surfaces, which represent the cantilever body and the bottom of the silicon substrate, respectively. We carefully check that the simulations are performed in a regime in which the results are independent of the distance between such boundaries and the sample surface. From the resulting steady-state temperature distribution (**Figure 2c-d**) we estimate the PTIR signal for the different values of the PMMA thickness. This can be done by computing the thermal expansion as $\Delta z = \alpha_{\text{th}} \int_0^{z_0} u(z) dz$, where $\alpha_{\text{th}}$ is the linear thermal expansion coefficient and $u(z)$ is temperature increase in the sample, i.e. $u(z) = T(z) - 300$ K. As demonstrated in **Figure 2e-f**, the thickness dependence of the PTIR signal retrieved by such simulations is superlinear for the Au tip and roughly linear for the Pd tip, which is qualitatively different from what we found in the experiments. The modeling framework needs therefore to be refined in order to reproduce the qualitative thickness dependence of the experimental findings.

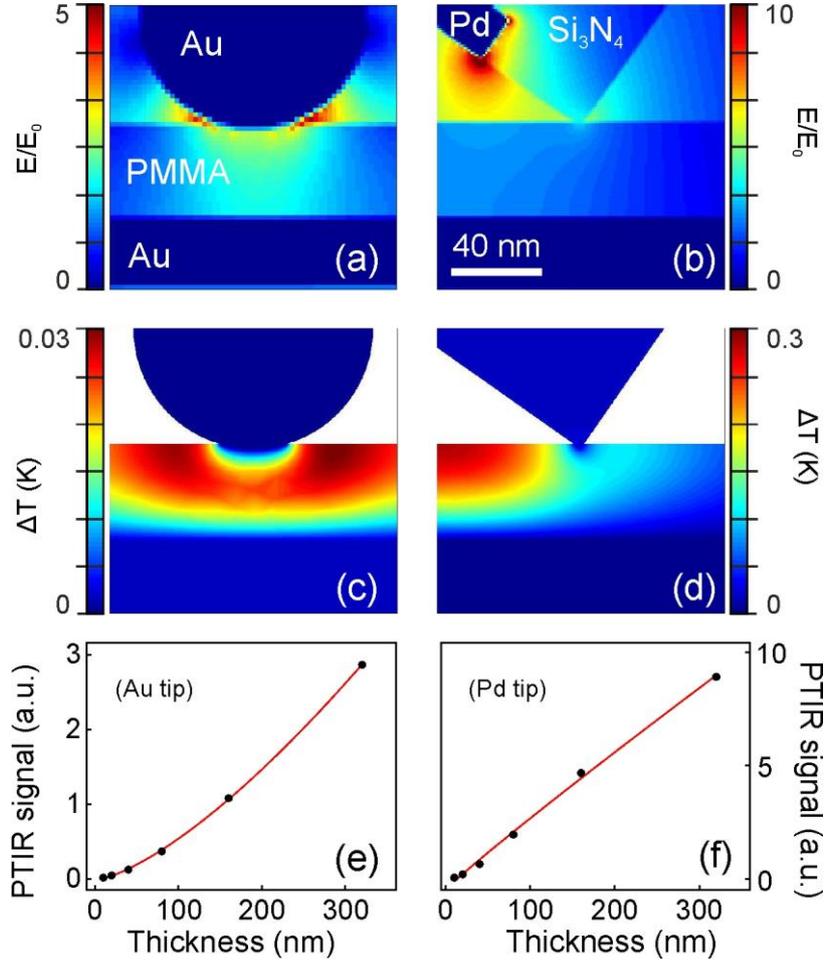

**Figure 2:** Combined optical and thermal simulations of the PTIR signal. (a)-(b) Simulated maps of the electric field magnitude for the Au and Pd tip, respectively. (c)-(d) Simulated maps of the steady-state temperature for the Au and Pd tip, respectively. (e)-(f) PTIR signal as a function of the thickness of the PMMA film, as calculated from the temperature maps for the Au and Pd tips, respectively. Solid lines are a

At this point, in order to further simplify our model and quickly evaluate the role of different boundary conditions and simulation parameters without employing lengthy numerical simulations, we also consider an approximate solution by solving the steady-state heat-diffusion equation in one dimension, along the normal to the film surface, which corresponds to neglecting in-plane temperature gradients in the sample volume right below the tip (see **Figure 2c-d**). We write the heat-diffusion equation inside the PMMA film in the form:

$$\rho C_\mathrm{p} \frac{\partial u(z,t)}{\partial t} = \eta \frac{\partial^2 u(z,t)}{\partial^2 z},$$

where $C_\mathrm{p}$ is the specific heat of PMMA, $u(z,t)$ is the temperature increase at the position $z$ and time $t$, $\eta$ is the thermal conductivity of the PMMA, and $Q(z)$ is the source term retrieved from the simulated optical absorption. In this simplified model, therefore, the only term depending on the choice of the tip is $Q(z)$, so the two different behaviours must be obtained when we change such source term. This is done by exctracting the electric field intensity along the thickness $z$ of the PMMA film from our optical simulations. In order to take the whole optical hot spot into account, we integrate the field intensity along the coordinate parallel to the sample surface, over a radius of 60 nm centered on the tip axis. Since the absorption by the tip and the Au film provides a negligible contribution to the temperature of the PMMA film (as verified with additional two-dimensional optical and thermal simulations), in the semi-analytical model we apply the heat sink boundary conditions directly at the PMMA/substrate ($z = 0$) and PMMA/air ($z = z_0$) interfaces. Under these assumptions, the boundary conditions at the two ends of the PMMA (defined by the coordinates 0 and $z_0$) become:

$$u(0) = u(z_0) = 0 \text{ K}.$$

The differential equation is then solved numerically obtaining results that are in very good agreement with the previous finite-element simulations (**Figure 3a-b**), with a superlinear behavior for the Au tip and a roughly linear behavior for the Pd tip. The analytical model can now be easily modified to simulate further mechanisms that might contribute to the results of the measurement. In particular, we consider the effect of thermal resistances at the tip/sample and sample/substrate interfaces. We then need to modify the semi-analytical model in order to explicitly account for their presence since they introduce a discontinuity in the temperature at the boundaries. The relation between the temperature discontinuity $\Delta u$ at one of the interfaces and the associated heat flux $j(z) = -\eta \frac{du}{dz}$ is given by $j(z) = \frac{\Delta u}{R_{\text{th}}}$. Therefore, the boundary conditions for the heat flux can be written in the following way:

$$j(0) = -\eta \frac{du(0)}{dz} = -\frac{u(0^+) - u(0^-)}{R_1} = -\frac{u(0^+)}{R_1},$$

$$j(z_0) = -\eta \frac{du(z_0)}{dz} = -\frac{u(z_0^+) - u(z_0^-)}{R_2} = -\frac{u(z_0^+)}{R_2},$$

in which we still consider that the tip and the substrate temperatures are unperturbed, which is represented by the condition:

$$u(0^-) = u(z_0^+) = 0 \text{ K}.$$

In general, the two interface thermal resistances of unit area $R_1$ and $R_2$ have different values. However, in this simplified analysis we will assume a symmetric situation, employing the same value $R$ for both of them. Moreover, since the interface under consideration is the one between PMMA and Au, we find ourselves in the unique position to directly extrapolate its experimental value from the recent publication by Chae *et al.* [8], already discussed in the introduction, in which the authors perform dynamic PTIR measurements and quantify the heat conductivity of a material. Their results show that for very thin samples the heat conductivity is different from the one measured in the bulk and is dependent on the value of the thickness. This can indeed be interpreted as due to the presence of interfacial thermal resistances that affect the temperature distribution in the film and consequently the PTIR signal. With a simple model, we are able to extrapolate a value for the interfacial thermal resistances of unit area directly from their results. The value that we employ is $R = 10^{-6} \text{ m}^2 \text{ K W}^{-1}$. It should be noted that, since we have no other available reference, we attribute the same value also to the interface thermal resistance between the silicon nitride tip and the PMMA film.

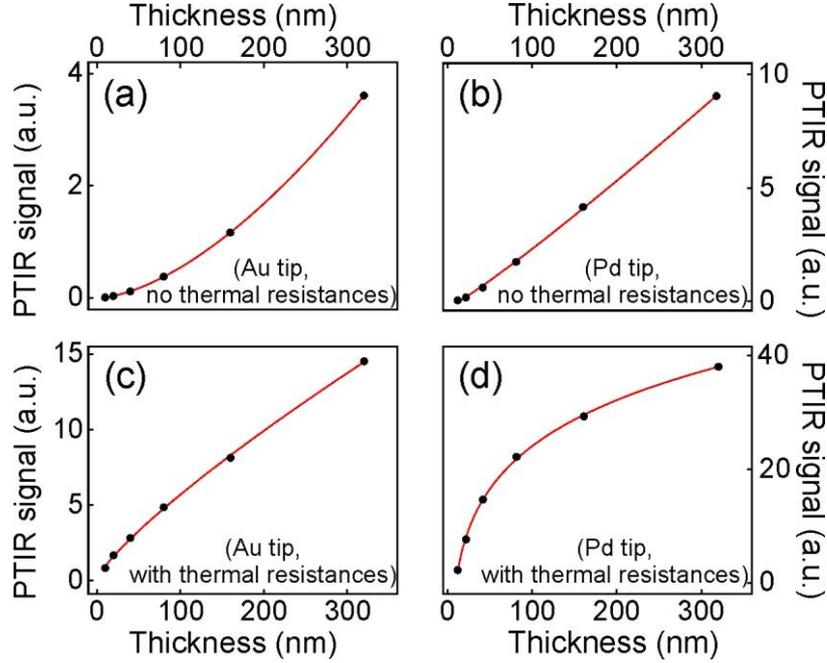

**Figure 3:** Results of the one-dimensional semi-analytical approach. (a)-(b) PTIR signals as a function of the film thickness for the Au and Pd tips, respectively, without any interface thermal resistance. (c)-(d) PTIR signals as a function of the film thickness with interface thermal resistances for the Au and Pd tips, respectively. Solid lines are a guide for the eye.

The results for the two tips are pictured in **Figure 3c-d**: we see a dramatic change in the qualitative behaviour of the PTIR signal for both tips because of the introduction of the interfacial resistances. The gold tip now shows a roughly linear trend, whereas the silicon nitride one determines a sublinear dependence of the PTIR signal on the PMMA thickness. We stress again that the reason for this difference between the two tips is purely due to the source term $Q(z)$, representing the optical absorption in the two cases, since it is the only parameter changing in the two calculations. This points out that the field localization capabilities of the tip also play a crucial role, as expected, in determining the thickness dependence of the PTIR signal: the larger the degree of local enhancement is, the more surface sensitive the PTIR signal is, with a more pronounced tendency to saturate for larger thicknesses. To further corroborate this argument, we plot in **Figure 4** the simulated $z$ profiles of the electric field intensity below the apex of the tip for both the Au (panel a) and Pd (panel

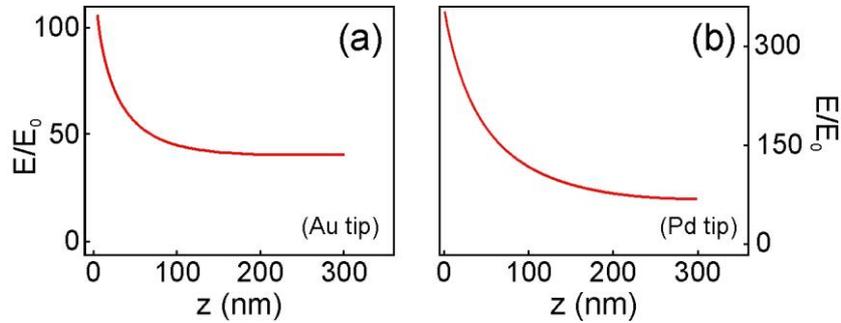

**Figure 4:** Simulated electric field intensity profiles along the normal to the sample surface, after integration over a radius of 60 nm around the tip axis, for the Au (panel a) and Pd (panel b) tip. The results are normalized to those for a plane wave in vacuum.

b) tip, demonstrating the higher degree of energy confinement granted by the resonant antenna for the Pd tip.

We can therefore conclude that the presence of interfacial thermal resistances influences the PTIR signal not only quantitatively but also qualitatively: the discontinuity at the interfaces changes both the steady-state temperature distribution and its thickness dependence. In order to further confirm these results, we employ the same idea in the two-dimensional finite-element thermal simulations. We model the thermal resistances as very thin layers of insulators and we minimize their thickness in order to reproduce an interfacial behaviour, i.e. create a discontinuity in the temperature distribution, while avoiding any significant changes in the simulated geometry. Eventually we obtain an object which replicates the boundary conditions applied in the semi-analytical one-dimensional model (keeping the same value of thermal resistance of unit area as mentioned before). We put such insulating layers at the tip/sample and sample/substrate interfaces. In this way, we effectively modify the boundary conditions applied to the PMMA film. The results of the simulations performed in this environment indeed show modified steady-state temperature distributions with respect to the case where no interface thermal impedances are considered (**Figure 5a-b**): the temperature hot spot is more spread along the directions parallel to the sample surface, while it is almost fully confined along the thickness of the PMMA film. These distributions give rise to different behaviours of the PTIR signal as a function of the film thickness: for the gold tip we observe an almost linear dependence, while for the silicon nitride tip a sublinear one (**Fig 5c-d**). This is in good agreement with the experimental results and with the semi-analytical model that we employed before. We also performed an analysis on the specific weight of the resistances: by considering just one of the two (with the other one turned off), we simulate the PTIR signal as before. By doing so, we observe that the resistance at the sample/substrate interface determines a more evident change in the dependence of such signal on the thickness of the sample.

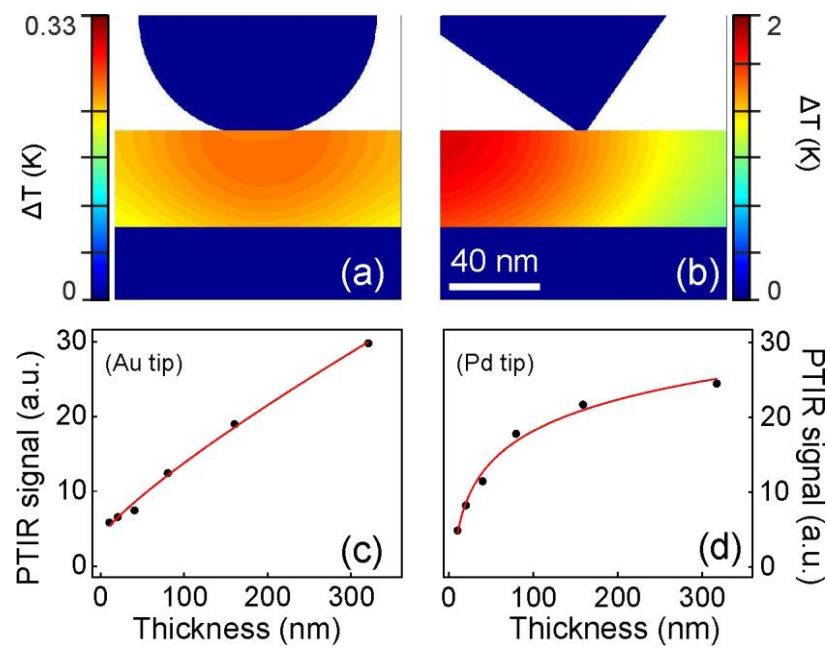

**Figure 5:** Results of the thermal simulations of the PTIR signal taking interface thermal resistances into account. (a)-(b) Simulated maps of the steady-state temperature for the Au and Pd tip, respectively. (e)-(f) PTIR signal as a function of the thickness of the PMMA film, as calculated from the temperature maps for the Au and Pd tips, respectively. Solid lines are a guide for the eye.

**Conclusions**

In conclusion, we have benchmarked the combined use of optical and thermal simulations against a systematic set of experimental results where the PTIR signal from the carbonyl line in a PMMA film is measured as a function of the film thickness with two different tip geometries. Our analysis demonstrates that any modelling of a PTIR experiment cannot neglect the crucial role played by interfacial thermal resistances. Moreover, the thickness dependence of the PTIR signal is also confirmed to be related to the electromagnetic field confinement properties of the employed tip, with a higher degree of confinement around the apex being associated with a weaker dependence on the film thickness. These results offer a guideline for the qualitative and quantitative interpretation of any PTIR analysis.